\documentclass[12pt]{iopart}

\usepackage{iopams}
\usepackage{setstack}
\usepackage{graphicx}
\usepackage{subeqn}
\usepackage{color}

\newcommand\dd{{\rm{d}}}

\newcommand\zR{{z_{\rm{R}}}}
\newcommand\dF{{d_{\rm{FWHM}}}}

\newcommand\uuD{\underline{\underline{\rm{D}}}}

\newcommand\uN{\underline{N}}
\newcommand\tN{\tilde{N}}
\newcommand\utN{\underline{\tilde{N}}}

\newcommand\uuT{\underline{\underline{\rm{T}}}}
\newcommand\uuM{\underline{\underline{\rm{M}}}}

\def\uuM{\mathbf{M}}
\def\uuT{\mathbf{T}}
\def\uuD{\tilde{\mathbf{M}}}
\def\uN{\mathbf{N}}
\def\utN{\tilde{\mathbf{N}}}

\begin{document}

\title[]{Multiple ionization of neon by soft X-rays at ultrahigh intensity}
\author{R Guichard$^{1,2}$\footnote{current address: roland.guichard@upmc.fr}, M Richter$^3$\footnote{current address: mathias.richter@ptb.de}, J-M Rost$^{1}$, U Saalmann$^{1}$, A A Sorokin$^{3,4,5}$ and K Tiedtke$^4$}
\address{$^1$Max-Planck-Institut f\"ur Physik komplexer Systeme, MPIPKS\\ 
N\"othnitzer Stra{\ss}e 38, 01187 Dresden, Germany}
\address{$^2$UPMC Universit\'e Paris 06, UMR 7614, Laboratoire de Chimie Physique Mati\`ere et Rayonnement, F-75005 Paris, France}
\address{$^3$Physikalisch-Technische Bundesanstalt, PTB \\
Abbestra\ss e 2-12, 10587 Berlin, Germany}
\address{$^4$Deutsches Elektronen-Synchrotron, DESY\\
Notkestra\ss e 85, 22603 Hamburg, Germany}
\address{$^5$Ioffe Physico-Technical Institute,\\
 Polytekhnicheskaya 26, 194021 St. Petersburg, Russia}

\date{\today}

\begin{abstract}
At the free-electron laser FLASH, multiple ionization of neon atoms was quantitatively investigated at 93.0 eV and 90.5 eV photon energy. For ion charge states up to 6+, we compare the respective absolute photoionization yields with results from a minimal model and an elaborate description. Both approaches are based on rate equations and take into acccout a Gaussian spatial intensity distribution of the laser beam. From the comparison we conclude, that photoionization up to a charge of 5+ can be described by the minimal model. For higher charges, the experimental ionization yields systematically exceed the elaborate rate based prediction.
\end{abstract}


\medskip

With the construction of X-ray lasers during the last years applying the free-electron laser (FEL) concept of self-amplified spontaneous emission (SASE) \cite{Ackermann, Shintake, Emma, David}, a number of experiments have become feasible which elucidate the principles of photon-matter interaction in a new parameter regime (\cite{Bostedt, Berrah} and refs. therein). The combination of high frequency and high intensity allowed in particular new insight into the photoelectric effect and the photoionization process \cite{Wabnitz, Siedschlag, Saalmann, Sorokin, Nagler, Young, Vinko, Richter,Rudek}. Even simple photoionization involves in a strict sense multi-electron dynamics as soon as the photoelectron is not ejected dominantly from the highest occupied orbital rendering a comprehensive theoretical description difficult. For more complex atomic targets such as xenon, multiphoton absorption in the energy range of the giant resonance poses in addition the question, which role collective electron excitations play in the high frequency -- high intensity domain \cite{Sorokin, Richter2}.  Due to these difficulties, the role of electron correlation for the understanding of multiphoton ionization in the X-ray regime has not been settled to date \cite{Richter2, Makris, Richter3, Lambropoulos, Lambropoulos2}. Hence, it is worthwhile to carefully assess in which cases simple processes, such as sequential photoionization, describe the experiments quantitatively and where deviations ask for explanations with a more involved dynamics.
In the VUV-regime it was found in an experiment with argon atoms that sequential ionization describes the measured  yields for different charge states well \cite{Motomura}.
 In the hard X-ray regime, it could be shown that sequences of single ionization in inner electron shells describe photon-matter interaction in the hard X-ray range and for light elements such as neon (Ne) well \cite{Young}.  The present work determines in a combined experimental-theoretical study, to which extend a sequential description for Ne is also valid in the EUV regime. For this purpose, results of ion spectroscopy on Ne atoms obtained in a focused beam of the free-electron laser FLASH in Hamburg \cite{Ackermann} at 93.0 eV and 90.5 eV photon energy are compared with a minimal analytical model for sequential ionization we have developed and with recent theoretical results from a more elaborated approach, including also non-sequential rates \cite{Lambropoulos2}. We find  that the  sequential minimal model describes the first five ionization stages reasonably well. The cross sections for higher charged ions, where sequential ionization can be excluded, are systematically underestimated by the elaborate rate description. Although the effect is weaker as compared to Xe with its giant 4$d$ resonance in the EUV \cite{Sorokin}, the results point to the importance of electron correlation for photoionization, also at higher photon frequencies and high intensity.

\begin{figure}[b]
\begin{center}
\includegraphics[width=0.7\linewidth]{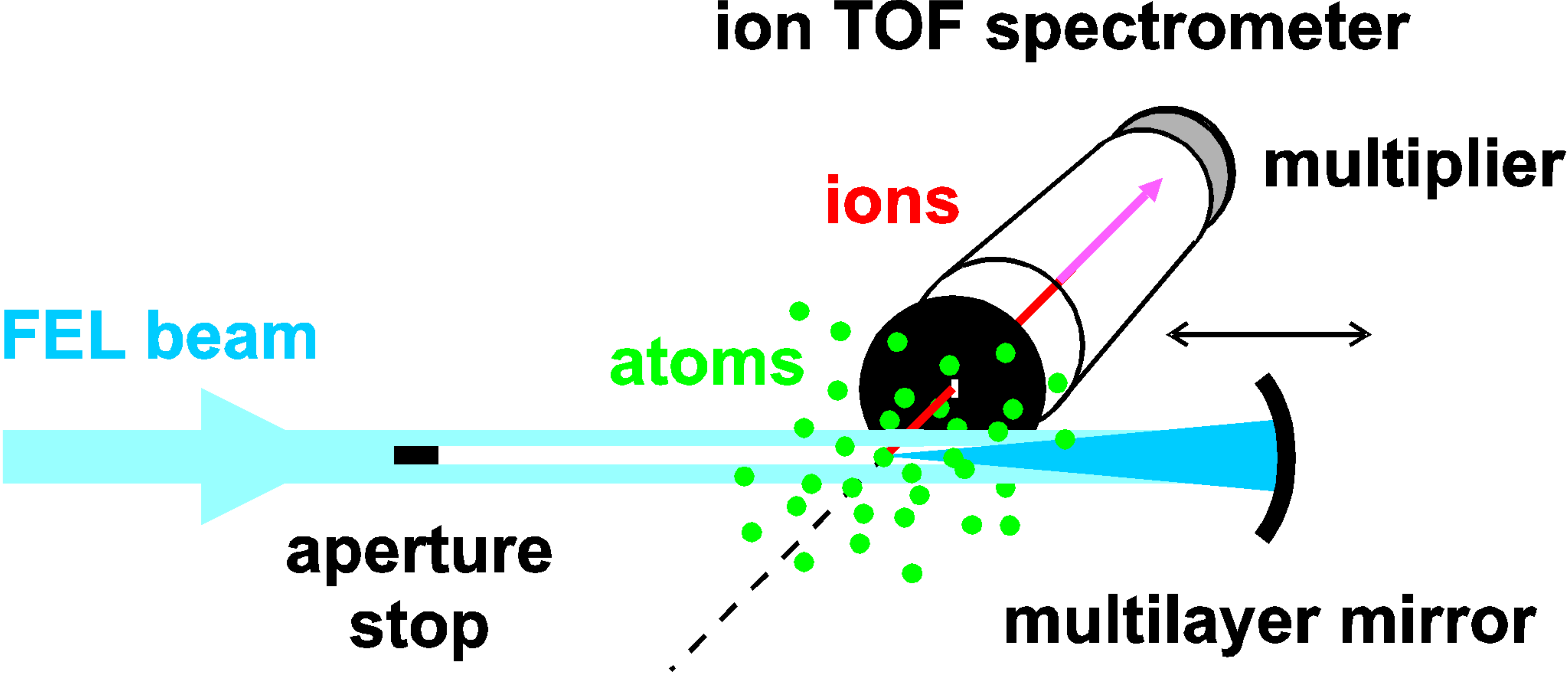}
\end{center}
\caption{Experimental setup for the investigation of atoms by ion time-of-flight (TOF) spectroscopy in a free-electron Laser (FEL) focus of a spherical narrow-bandwidth multilayer mirror. \label{fig1}}
\end{figure}

The experiments were carried out by applying ion time-of-flight (TOF) spectroscopy as shown in Fig.~\ref{fig1} \cite{Sorokin, Richter3, Sorokin2}. The FEL beam was focused by the use of a spherical Si/Mo narrow-bandwidth multilayer mirror under normal incidence with a focal length of 20 cm and a reflectance between 60\% and 70\% \cite{Feigl}, depending on the photon energy and confirmed with a relative standard uncertainty of 10\%. The minimum focus diameter as small as $\dF$ = (2.6 $\pm$ 0.5) $\mu$m was derived from the target depletion effect \cite{Sorokin2}. The absolute FEL pulse energy in the $\mu$J regime was monitored with a relative standard uncertainty of 15\% on a shot-to-shot basis by means of calibrated gas-monitor detectors \cite{Tiedtke}. The mirror could be moved along the FEL beam in order to shift the focus in back-reflection geometry into and out of the interaction volume of our ion TOF spectrometer to vary the effective FEL beam cross section \cite{Sorokin, Richter}. As a result, the peak intensity (irradiance) of the FEL pulses with pulse durations of $\Delta t_{\rm{FWHM}}$ = (15 $\pm$ 5) fs \cite{Richter2} could be varied from 10$^{13}$ to 4$\times$10$^{15}$ W/cm$^2$ and measured with a standard uncertainty of 42\%. Ne filled the experimental vacuum chamber homogeneously at the low pressure of about $10^{-4}$ Pa to avoid any interaction between neighboring atoms and ions. The pressure was measured by a calibrated spinning rotor gauge and the temperature by a calibrated Pt100 resistance thermometer. The homogeneous electric extraction field was sufficiently high to collect and register all ions generated within the interaction volume of the TOF spectrometer by an open electron multiplier operated in the analog mode, i.e., by measuring the charge accumulated on the multiplier anode. The entrance aperture of the spectrometer had an extension of 350 $\mu$m in the beam direction and 1 mm perpendicular to the FEL beam. In front of the experimental chamber, a horizontal beam stop of a 1.5 mm in height was introduced which enabled us to diaphragm ions produced by the incident unfocused beam (cf. Fig.~\ref{fig1}). TOF spectra were averaged over typically 500 consecutive FEL shots with pulse-to-pulse intensity fluctuations varying from 35\% to 45\%.

\begin{figure}[h]
\begin{center}
\includegraphics[width=0.7\linewidth]{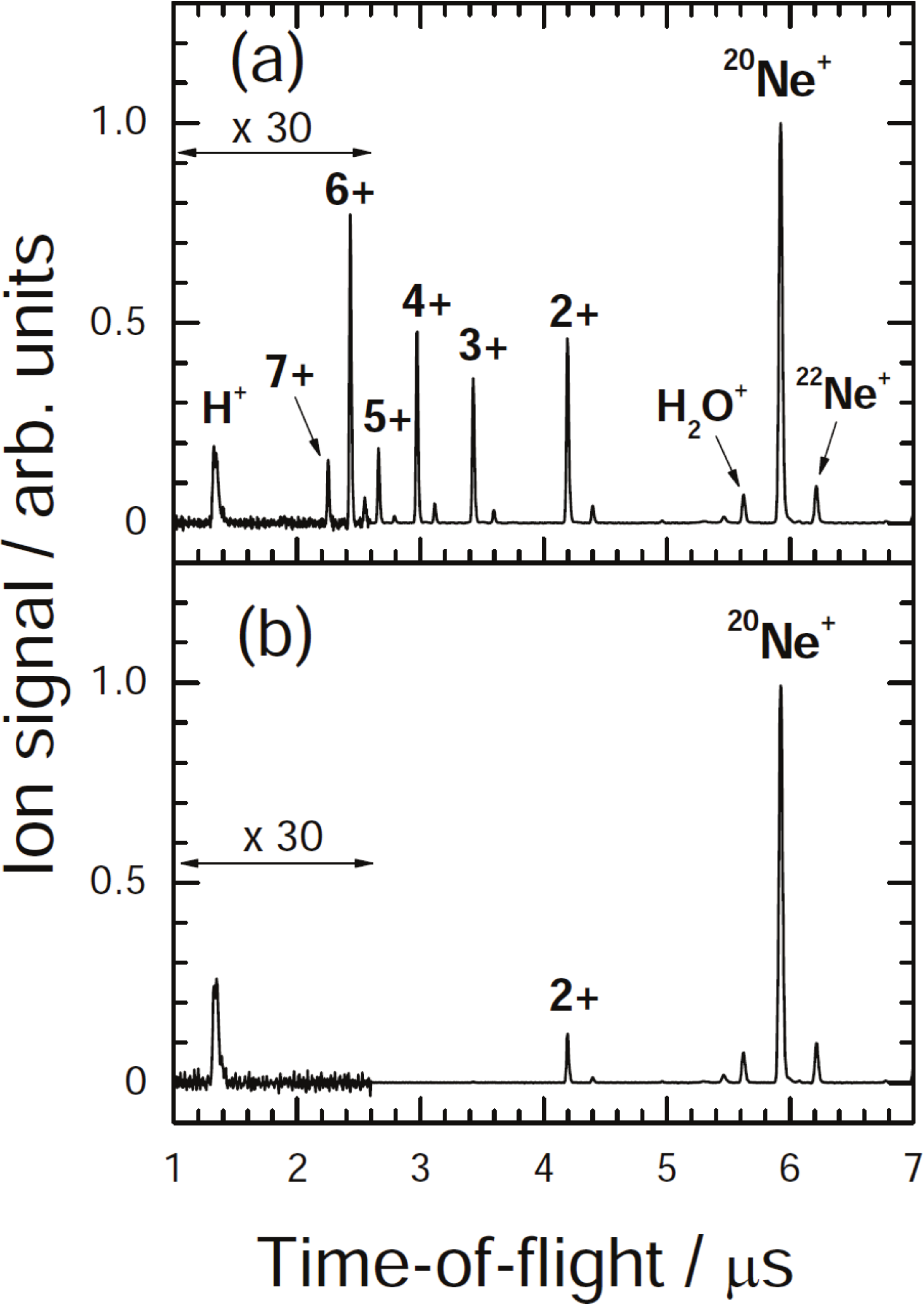}
\caption{Ion time-of-flight (TOF) spectra of Ne taken at the photon energy of 93.0 eV and the peak intensity of (a) 4$\times$10$^{15}$ W/cm$^2$ and (b) 2$\times$10$^{13}$ W/cm$^2$. In the TOF regime below 2.6 $\mu$s, the ion intensities were multiplied by a factor of 30.\label{fig2}}
\end{center}
\end{figure}

\begin{figure}[h]
\begin{center}
\includegraphics[width=0.7\linewidth,angle=270]{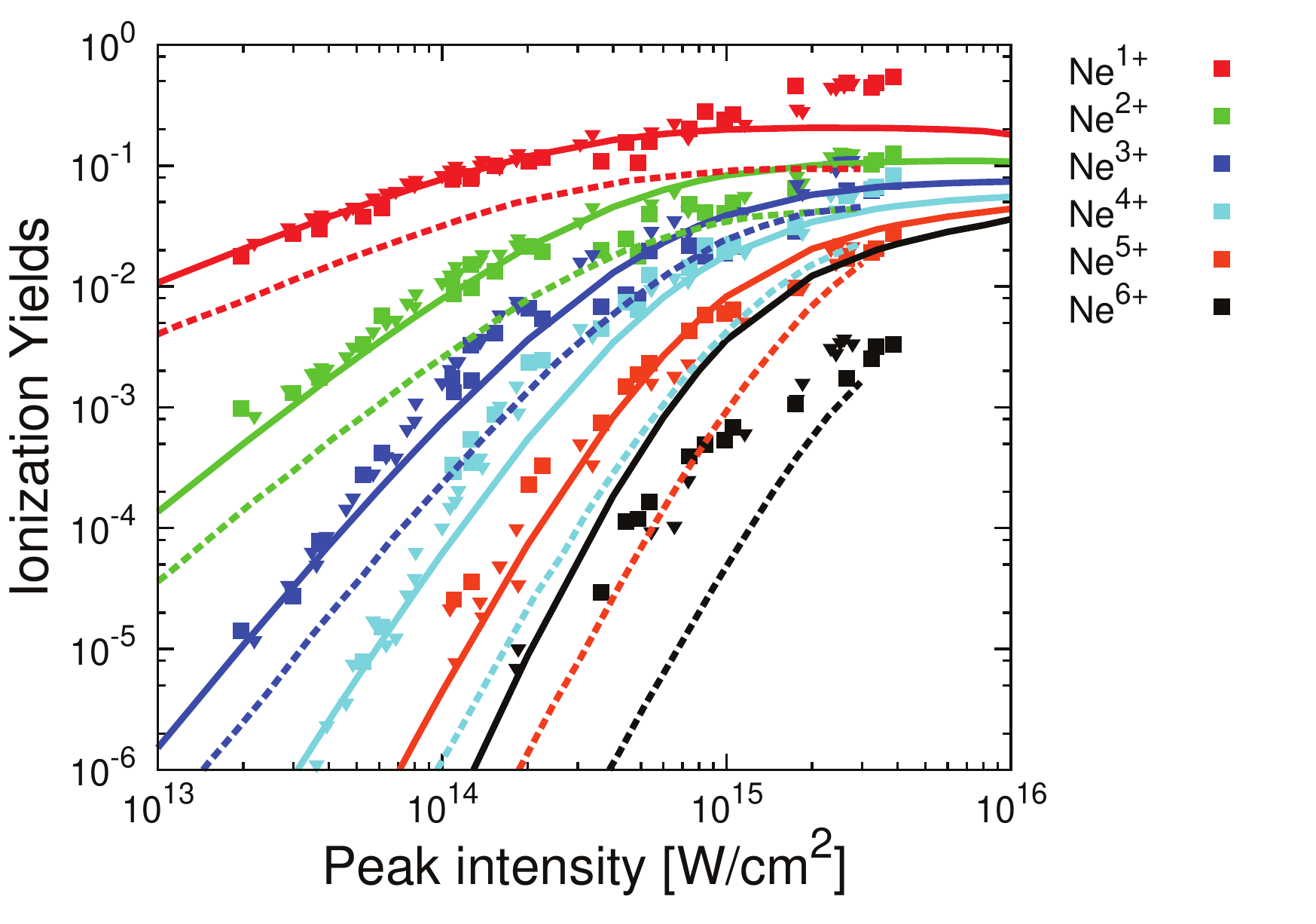}
\caption{Experimental data for Ne ionization yields measured at 93.0 eV (\fullsquare) and 90.5 eV (\fulltriangledown) photon energy with pulses of 15 fs duration compared with results from the present minimal model (\full) and results from the elaborate rate description for 30 fs pulse length  of Ref. \cite{Lambropoulos2} (\dashed). \label{fig3}}
\end{center}
\end{figure}

Fig.~\ref{fig2} shows typical ion TOF spectra of Ne taken at the highest and lowest intensity level, corresponding to the regime of multi- and single-photon excitation, respectively. Charged states up to 7+ were observed. Up to 6+, i.e., up to the complete removal of the 2$p$ electron shell, the corresponding ionization yields were deduced from the ion signals by normalization to the absolute ion detection efficiency and to the number of target atoms within the interaction volume. The absolute detection efficiencies for singly charged ions were directly measured at low intensity levels where single-photon excitation is dominating, using well-known photoionization cross section data of Ne (\cite{Tiedtke} and refs. therein). The detection efficiencies for higher charge states were deduced from those of singly charged ions assuming that the efficiency is proportional to the ion impact velocity \cite{Schram, Stockli}. The number of target atoms was calculated from the interaction volume and the atomic particle density $n_a$. The latter was determined by $n_a = p/kT$, where $p$ is the target gas pressure, $T$ is the temperature, and $k$ is the Boltzmann constant. The interaction volume was calculated taking into account the minimum focus diameter, the distance of the interaction volume of the TOF spectrometer from the focus, the width of the spectrometer entrance aperture, and the focused beam divergence. The latter was obtained using the known focal length of the mirror and measuring the size of the unfocused beam incident on the mirror. The relative standard uncertainty for the ionization yield is estimated to amount to 36\% which arises from the uncertainties for target density (2\%), interaction volume (30\%), and detection efficiency (20\%). Fig.~\ref{fig3} shows the experimental ionization yields for the Ne charge states from 1+ to 6+ as a function of the peak intensity, with filled and open symbols for measurements at 93.0 eV and 90.5 eV photon energy, respectively. 

The dependence of the different ionization yields on the peak intensity was also calculated by means of a minimal analytical model which is based on the stepwise removal of electrons by the sequential absorption of photons, i.e., Ne$\overset{\hbar\omega}{\rightarrow}$ Ne$^{+}$ $\overset{\hbar\omega}{\rightarrow}$ \dots\ $\overset{\hbar\omega}{\rightarrow}$ Ne$^{n+}$, through a set of rate equations \cite{Lambropoulos3}

\begin{subequations}\label{seq}
\begin{eqnarray}
\dot{N}_0(t)&=&-\sigma_{01}F(t)N_0(t),\\
\dot{N}_1(t)&=&\sigma_{01}F(t)N_0(t)-\sigma_{12}F(t)N_1(t),\\
\dot{N}_2(t)&=&\sigma_{12}F(t)N_1(t)-\sigma_{23}F(t)N_2(t),\\
&\vdots&\nonumber
\end{eqnarray}
\end{subequations}

\noindent with $F(t) = F_0 f(t)$ where $F_0$ is the peak photon flux per unit area and $f(t)$ the dimensionless temporal shape of the photon pulses. $\sigma_{jk}$ denotes the single photon ionization cross section from charge state $j$ to $k$. With the new time variable
\begin{equation}
\label{pulselength}
\tau(t) = \int^t_{-\infty} \dd t'\, f(t')\,,
\end{equation}
 which corresponds for infinite time to an effective pulse length $\tau_{\infty}=\tau(t{\to}\infty)$, the set of Eqs.~\ref{seq} may be rewritten as:

\begin{equation}
\frac{d\uN(I,\tau)}{d\tau}=\uuM\;\uN(I,\tau)\,,\label{System}
\end{equation}

\noindent with the \emph{time-independent} matrix 

\begin{equation}\label{eq:Mmat}
\uuM=
\frac{I}{\hbar\omega}\left(
\begin{tabular}{@{}ccccc}
$-\sigma_{01}$&0&$\dots$&0&0\\
$+\sigma_{01}$&$-\sigma_{12}$&$\dots$&0&0\\
0&$+\sigma_{12}$&$\dots$&0&0\\
$\vdots$&$\vdots$&&$\vdots$&$\vdots$\\
0&0&$\dots$&$+\sigma_{n-1,n}$&0
\end{tabular}
\right).
\end{equation}

\noindent and ${\uN} (I,\tau)\equiv \{N_0(I,\tau),N_1(I,\tau),N_{2}(I,\tau),\ldots\}^\dagger$ where the photon flux $F_{0}$ is replaced by the peak intensity $I=F_{0}\,\hbar\omega$. The formal solution now reads as:

\begin{equation}
\uN(I,\tau)=\exp\left(\uuM\tau\right)\uN(0)\,.\label{Solution}
\end{equation}

With the initial condition $\uN(\tau{=}0) \equiv \uN(t{\to}-\infty) = \{1,0,0,\dots\}^\dagger$, the vector components $N_{n>0} (I, \tau)$ may be interpreted as the respective ionization probabilities while $N_0 (I, \tau)$ is the remaining fraction of neutral atoms.

\noindent From this general sequential ionization description we arrive at our minimal model by using identical cross sections $\sigma_{jk} = \sigma$ for all $jk$, i.e., we need only a single external parameter $\sigma$. For the first ionization steps, the use of identical step-wise cross sections can be justified  since $\sigma_{01}$, $\sigma_{12}$, $\sigma_{23}$, and $\sigma_{34}$ available in the 90 to 95 eV photon energy range, do not vary by more than 12\% from their mean value of about $4.2\times 10^{-18}$ cm$^2$ as can be seen in Tab. \ref{tab1}. 

\begin{table}
\caption{Single-photon cross sections $\sigma_{jk}$ in units of $10^{-18}$\,cm$^2$,
from experiment \cite{Tiedtke,Marr} (bold), using the  GIPPER code \cite{gipper} (italic)  and a fitting procedure \cite{Verner}.}
\centerline{\begin{tabular}{cccc}
  $\sigma_{01}$ & $\sigma_{12}$ & $\sigma_{23}$ & $\sigma_{34}$\\ \hline
 \textbf{4.5} & \textit{4.07} & \textit{3.98} & \textit{3.71}\\
 & 4.31 & 4.55 & 3.96\\ \hline
 \end{tabular}}
\label{tab1}
\end{table} 

\noindent In this minimal model one can solve (\ref{System}) analytically with the explicit expression 

\begin{equation}
N_{n}(I,\tau_{\infty})=\frac{\lambda^n}{n!}\exp\left(-\lambda\right),
\qquad\lambda= \sigma I\tau_{\infty}/\hbar \omega.
\label{Seqeq2}
\end{equation}

\bigskip


For comparison with the experimental results, the theoretical data has to be convoluted with the spatial intensity distribution of the focused FEL beam within the interaction volume. We have chosen


\begin{equation}
I(I_0,r,z)=\frac{I_0}{1+\left(z/\zR \right)^2}\exp\left(-\frac{2r^{2}}{w_0^{2}\left[1+\left(z/\zR \right)^2\right]}\right)\,,\label{GB}
\end{equation}

\noindent i.e., a Gaussian distribution in radial direction $r$. Its standard deviation increases with $z$ along the beam from the value $w_0=\dF /2\sqrt{\ln2/2}=2.208$\,$\mu$m at $z=0$ while $\zR$ = 0.13 mm denotes the estimated Rayleigh length of the focus for our experiments. The results of the minimal model according to Eq.~(\ref{Seqeq2}) (with $\sigma = 4.2\times 10^{-18}$ cm$^2$, $\tau_{\infty}=\sqrt{\pi/4\ln 2}\,\Delta t_{\rm{FWHM}}=16$ fs, $\hbar\omega=(93.0+90.5)/2$ eV) convoluted with the spatial intensity distribution of Eq.~(\ref{GB}) are shown as a function of the peak intensity $I$ with solid lines in Fig.~\ref{fig3}. For comparison, also the results of the elaborate rate description of the Ne ionization yields are shown by dashed lines \cite{Lambropoulos2}.


As can be seen in Fig.~\ref{fig3}, the experimental ion yields from Ne$^{1+}$ to Ne$^{5+}$ are well reproduced by our minimal analytical model with the parameter $\lambda$ from Eq.~(\ref{Seqeq2}) globally scaled by a factor 1.20 within the experimental uncertainties for $\lambda$ of about 26\%. The agreement between experimental data and minimal model within the experimental uncertainties demonstrates both, the sequential character of the first few ionization steps, and that the photon intensity distribution of Eq.~(\ref{GB}) is appropriate for the spatial averaging. The discrepancies between experiment and minimal model at high intensity ($>10^{15}$ W/cm$^2$) might be explained by out-of-focus radiation, due to spherical aberrations, hitting the neutral atoms which are continuously distributed within the vacuum chamber. The predictions of the elaborate rate description from Ref.~\cite{Lambropoulos2} for the Ne$^{1+}$, Ne$^{2+}$, and Ne$^{3+}$ yields are in the same order of magnitude, although a bit lower which is most likely due to the fact that the laser pulse length is with 30 fs longer than in the experiment (16 fs) but may also be due to the different way of spatial averaging. The yield for Ne$^{4+}$ and higher ionization from Ref.\cite{Lambropoulos2}, on the other hand, is significantly lower than the values of our minimal analytical model and the experiment. Note, that the threshold energy for the Ne$^{3+}$ $\rightarrow$ Ne$^{4+}$ ionization process amounting to about 97 eV \cite{Kelly} and, hence, exceeds our photon energies of 93.0 eV or 90.5 eV, respectively. As a consequence, this ionization step starting from Ne$^{3+}$ requires at least two photons, so that rendering its strength much weaker. In our minimal analytical model, sequential single-photon ionization is assumed but without explicit reference to the ionization thresholds. Interestingly, the yields for Ne$^{4+}$ and Ne$^{5+}$, which cannot be reached by fully sequential photo ionization, are nevertheless well reproduced by the minimal model. Starting from Ne$^{6+}$, the minimal model produces unrealistically large yields.

On the other hand, the data sets in Fig.~\ref{fig3} clearly demonstrate that the predictions of the elaborate rate model from \cite{Lambropoulos2} are too low compared to the experimental ionization yields for the higher charges Ne$^{4+}$, Ne$^{5+}$, and Ne$^{6+}$. These discrepancies reveal the limitations of rate based approaches, similarly as in the multiphoton ionization of Xe in the same photon energy range \cite{Sorokin, Richter3}. 


In summary, we have demonstrated that sequential ionization schemes describe the generation of lower charge states in multiphoton ionization by soft X-rays well. For the higher charges, our experimental ionization yields of Ne are systematically underestimated by the elaborate rate description, as already known from previous results for Xe. One may conclude that in nonlinear multiphoton-matter interaction electron correlation can be significant and should be taken into account to interpret experiments at the new X-ray laser facilities. We thank the FLASH team for a very successful FEL operation. Support from the Deutsche Forschungsgemeinschaft (DFG RI 804/5-1) is gratefully acknowledged.

\appendix

\section*{Appendix}

\section{Analytical solution to the coupled rate equations for sequential ionization}
\label{Annex}

In the general case the matrix $\uuM$  in (\ref{Solution}) reads
\begin{equation}
\uuM=
\left(
\begin{tabular}{@{}ccccc}
$-s_0$&0&$\dots$&0&0\\
$+s_0$&$-s_1$&$\dots$&0&0\\
0&$+s_1$&$\dots$&0&0\\
$\vdots$&$\vdots$&&$\vdots$&$\vdots$\\
0&0&$\dots$&$+s_{m-1}$&0
\end{tabular}
\right)
\end{equation}
where the $s_i = \sigma_{i,i+1}F_0$ are the single-photon absorption rates
for taking the ion from the charge state $ i $ to $ i+1 $. For simplicity we have assumed 
a maximal charge state  $m$. 
Thus, with the neutral fraction $N_{0}$, the matrix $\uuM$ has the size $(m{+}1)\times(m{+}1)$. 
After some algebra, one can find a transformation to the diagonal form
\begin{subequations}
\begin{eqnarray}
\uuD&=&\uuT^{-1} \uuM\, \uuT
\\
&=& \left(
\begin{tabular}{@{}ccccc}
$-s_0$&0&$\dots$&0&0\\
0&$-s_1$&$\dots$&0&0\\
$\vdots$&$\vdots$&&$\vdots$&$\vdots$\\
0&0&$\dots$&$-s_{m-1}$&0\\
0&0&$\dots$&0&0
\end{tabular}
\right)
\end{eqnarray}
\end{subequations}
with the transformation matrix
\def\PP#1#2#3#4{P_{#1}^{#2}/Q_{#3}^{#4}}
\def\m{m-1}\def\mm{m-2}
\begin{eqnarray}
&\uuT =&
\left(
\begin{array}{@{}cccccc}
1&0&\dots&0&0&0\\
\PP{0}{0}{0}{0}&1&\dots&0&0&0\\
\PP{0}{1}{0}{1}&\PP{1}{1}{1}{1}&\dots&0&0&0\\
\vdots&\vdots&&\vdots&\vdots\\
\PP{0}{\mm}{0}{\mm}&\PP{1}{\mm}{1}{\mm}&\dots&\PP{\mm}{\mm}{\mm}{\mm}&1&0\\
-\PP{1}{\m}{0}{\mm}&-\PP{2}{\m}{1}{\mm}&\dots&-\PP{\m}{\m}{\mm}{\mm}&-1&1
\end{array}
\right)
\end{eqnarray}
whereby the first two of the following products have been used 
\begin{subequations}
\begin{eqnarray}
&P_{k}^{l}\equiv\prod_{i=k}^{l}{s_i}
&Q_{k}^{l}\equiv\prod_{i=k}^{l}(s_{i+1}{-}s_k),\\
&R_{k}^{l}\equiv\prod_{i=k}^{l}(s_{i-1}{-}s_l)\qquad
&_{j}S_{k}^{l}\equiv\prod_{i=k}^{l}(s_{i}{-}s_j).
\end{eqnarray}
\end{subequations}
The latter two will be used below.
Note that $P_{k}^{l}=Q_{k}^{l}=R_{k}^{l}=1$ for $k>l$.
The denominators in the matrix elements of $\uuT$ require that $s_0\neq s_1 \neq s_2 \neq \dots \neq s_{m-1}$. 
The time evolution of the transformed system $\utN(\tau)=\uuT^{-1}\uN(\tau)$ is simply given by
\begin{equation}
\utN(\tau)=\exp\left(\tau \uuD \right)\utN(0)\,.\label{TS2}
\end{equation}
In general the initial condition is given by $\utN(0)=\uuT^{-1}\uN(0)$.
For the particular form $\uN(0)=\{1,0,0,\ldots\}^{\dagger}$,
it simplifies to
\begin{subequations}
\begin{eqnarray}
\tN_{n<m}(0) &=& P_{0}^{n-1}/R_{1}^{n}\\
\tN_{m}(0) &=& 1.
\end{eqnarray}
\end{subequations}
Finally, we can transform back to the original system by $\uN(\tau)=\uuT\utN(\tau)$
and obtain
\begin{subequations}\label{FT2}
\begin{eqnarray}
N_{n<m}(\tau)&=&\tN_n(\tau)
+\sum_{j=0}^{n-1}\frac{P_{j}^{n-1}}{Q_{j}^{n}}\tN_j(\tau)\nonumber\\
&=&P_{0}^{n-1}
\sum_{j=0}^{n}\frac{\exp\left(-s_{j}\tau\right)}{_{j}S_{0}^{n}}\label{eq:Nntau}\\
N_m(\tau)&=&1-\sum_{j=0}^{m-1} N_{j}(\tau),\label{eq:Nmtau}
\end{eqnarray}
\end{subequations}
whereby we have used $P_{0}^{n-1}=P_{0}^{j-1}P_{j}^{n-1}$ and $_{j}S_{0}^{n}=R_{1}^{j}Q_{j}^{n}$
in Eq.\,(\ref{eq:Nntau}) and $\sum_{n}N_{n}(\tau)\equiv1$ in Eq.\,(\ref{eq:Nmtau}).
The two Eqs.\,(\ref{FT2}) give the time-dependent probabilities $N_{n}$ for the various charge states $n$.
  
For the minimal model discussed in the text we need the expressions for degenerate absorption rates $s_0=s_1 =s_2 =\dots =s$.
These can be obtained by subsequent limits $s_{0}\to s$, $s_{1}\to s$, etc.\
of Eq.\,(\ref{eq:Nntau}).
One obtains the compact form 
\begin{eqnarray}\label{Seqeq}
N_{n<m}(\tau)&=&\frac{(s\tau)^n}{n!}\exp\left(-s\tau\right)
\end{eqnarray}
and $N_{m}(\tau)$ as in Eq.\,(\ref{eq:Nmtau}).

\section*{References}

\bibliographystyle{unsrt}


\end{document}